\documentclass[nofootinbib]{revtex4}
\pdfoutput=1
\usepackage[latin9]{inputenc}
\usepackage{color,graphicx,epsfig}
\usepackage{ifpdf}
\usepackage{amsmath}
\usepackage{bm}
\usepackage{color}
\usepackage[english]{babel}
\usepackage{graphicx}%
\usepackage{amsfonts}%
\usepackage{amssymb}
\usepackage{braket}
\usepackage{hyperref}

\definecolor{nicered}{rgb}{0.7,0.1,0.1}
\definecolor{nicegreen}{rgb}{0.1,0.5,0.1}
\hypersetup{colorlinks,citecolor= nicegreen,linkcolor= nicered}

\newcommand{\Green}[1]{{\color{nicegreen}{#1}}}

\makeatletter

\DeclareFontEncoding{LGR}{}{}
\DeclareTextSymbol{\~}{LGR}{126}
\newcommand{\lyxmathsym}[1]{\ifmmode\begingroup\def\b@ld{bold}
  \text{\ifx\math@version\b@ld\bfseries\fi#1}\endgroup\else#1\fi}

\providecommand{\tabularnewline}{\\}

\@ifundefined{textcolor}{}
{%
 \definecolor{BLACK}{gray}{0}
 \definecolor{WHITE}{gray}{1}
 \definecolor{RED}{rgb}{1,0,0}
 \definecolor{GREEN}{rgb}{0,1,0}
 \definecolor{BLUE}{rgb}{0,0,1}
 \definecolor{CYAN}{cmyk}{1,0,0,0}
 \definecolor{MAGENTA}{cmyk}{0,1,0,0}
 \definecolor{YELLOW}{cmyk}{0,0,1,0}
}

\makeatother

\begin{document}

\title{Higgs Uncovering Light Scalar Remnants of High Scale Matter Unification}

\author{Ilja Dor\v sner} \email[Electronic address:]{ilja.dorsner@ijs.si}
\affiliation{Department of Physics, University of Sarajevo, Zmaja od Bosne 33-35, 71000
  Sarajevo, Bosnia and Herzegovina}
  \affiliation{J. Stefan Institute, Jamova 39, P. O. Box 3000, 1001
  Ljubljana, Slovenia}

\author{Svjetlana Fajfer} \email[Electronic
address:]{svjetlana.fajfer@ijs.si} 
\affiliation{Department of Physics,
  University of Ljubljana, Jadranska 19, 1000 Ljubljana, Slovenia}
\affiliation{J. Stefan Institute, Jamova 39, P. O. Box 3000, 1001
  Ljubljana, Slovenia}

\author{Admir Greljo} 
\email[Electronic address:]{admir.greljo@ijs.si} 
\affiliation{J. Stefan Institute, Jamova 39, P. O. Box 3000, 1001
  Ljubljana, Slovenia}

\author{Jernej F. Kamenik} 
\email[Electronic address:]{jernej.kamenik@ijs.si} 
\affiliation{J. Stefan Institute, Jamova 39, P. O. Box 3000, 1001
  Ljubljana, Slovenia}
\affiliation{Department of Physics,
  University of Ljubljana, Jadranska 19, 1000 Ljubljana, Slovenia}

\begin{abstract}
We consider the impact of colored scalars that can couple directly to matter fields on the recently measured $h\to\gamma\gamma$ excess. Among all possible candidates only scalar states transforming as $(\mathbf{8},\mathbf{2},1/2)$ and $(\overline{\mathbf{6}},\mathbf{3},-1/3)$ under the Standard Model gauge group can individually accommodate the excess and remain in agreement with all available data. Current experimental constraints require such colored states to have an order one coupling to the Standard Model Higgs and a mass below $300$\,GeV. 
We use the best fit values to predict the correlated effect in $h\to Z\gamma$ and di-Higgs production. We furthermore discuss where and how these states appear in extensions of the Standard Model with primary focus on scenarios of matter unification. We revisit two simple $SU(5)$ setups to show that these two full-fledged models not only accommodate a light color octet state but correlate its mass with observable partial proton decay lifetimes.
\end{abstract}

\maketitle
\section{Introduction}
Colored scalars are frequent harbingers of new physics. For example, they are inherent to any theory of matter unification~\cite{Pati:1974yy}. Phenomenologically, colored scalars that can couple directly to matter fields are of particular interest. This feature makes them very appealing candidates for collider physics and precision flavor studies. For example, some have been suggested as possible explanations of the enhanced forward-backward asymmetry in $t\bar t$ production as measured at the Tevatron~\cite{ttbar} or the $(g-2)_\mu$ anomaly~\cite{g2}. More recently, it has been pointed out that some could also contribute significantly to the unexpectedly large CP asymmetry in the decays $D^0 \to K^+ K^-,\pi^+\pi^-$~\cite{Altmannshofer}.
Moreover, in a framework of simple one-particle extensions of the Standard Model (SM), these states can establish a unique link between collider and Planck scale physics via proton decay~\cite{Barr:2012xb}. They might even help induce tiny neutrino masses through loop effects~\cite{FileviezPerez:2009ud}.

It has been demonstrated that light colored scalars help improve unification of gauge couplings in the Grand Unified Theory (GUT) setting~\cite{Dorsner:2005fq,Dorsner:2006dj,Dorsner:2007fy,Perez:2008ry,Dorsner:2009mq,Stone:2011dn,Bertolini:2012im,Babu:2012iv,Babu:2012pp,Patel:2011eh}. It could thus be possible to establish a firm connection between the colored scalar mass scale and observable proton decay signatures in the minimal matter unification scenarios. Moreover, the colored states often reside in the same representation of the GUT group as the fields responsible for the electroweak symmetry breaking (EWSB). This could allow one to infer the strength of Yukawa couplings of colored states to matter. These, on the other hand, are subject to nontrivial constraints originating from low-energy flavor phenomenology and the requirement of viable fermion masses. What one can end up with is a highly predictive class of simple models that connect matter stability to low-energy phenomenology~\cite{Dorsner:2010cu,Dorsner:2011ai}.   

A generic feature of massive scalars in extensions of the SM is that they couple to the SM Higgs boson. Above the EWSB scale, one can write marginal---the so-called ``Higgs portal"~\cite{Patt:2006fw}---operators of the form $ \Phi^\dagger \Phi H^\dagger H$, where $H$ is the SM Higgs doublet, $\Phi$ is the new scalar weak (and color) multiplet and all possible color and weak contractions are assumed. After EWSB this will induce corrections to the masses of weak $\Phi$ components ($\Phi_i$) and couplings of the form $\Phi_i^\dagger\Phi_i h$, where $h$ is the physical Higgs boson. At the one-loop level, the presence of such interactions can affect Higgs production and decays as measured at the LHC~\cite{Chang:2012ta}. In fact preliminary data on the Higgs-like resonance with a mass of $m_h\simeq 125$\,GeV show an excess of events in the loop-induced $h\to\gamma\gamma$ decay channel with respect to the SM prediction at the $2\sigma$ level, while other, tree-level dominated Higgs decays to vector bosons show good consistency with the SM expectations~\cite{Espinosa:2012im, Giardino:2012dp,Corbett:2012dm}. If it were to persist at larger significance, such a deviation would point towards the existence of new degrees of freedom at the EW scale that couple to the Higgs boson.

Motivated by these observations, in the present study, we focus on the contributions of colored scalars to loop-induced production and decay channels of the Higgs boson through interactions of the form $ \Phi^\dagger \Phi H^\dagger H$ (a similar analysis based on earlier experimental results can be found in~\cite{Batell:2011pz}). In particular, we study fields with direct couplings to the SM matter fields that can thus play an interesting role in other collider and flavor observables. We single out two fields which accommodate the observed $h\to\gamma\gamma$ enhancement while remaining in good agreement with the other measured decay channels. Interestingly enough, these fields turn out to be sextets and octets of color. This precludes them from having leptoquark-like couplings to the ordinary matter.

We then proceed to show how and where these particular states appear in simple scenarios of matter unification. We revisit gauge coupling unification in two full-fledged models based on the $SU(5)$ gauge group~\cite{Georgi:1974sy} to demonstrate the connection between the lightness of these states and observable proton decay signatures. One of the models can also address the  $(g-2)_\mu$ anomaly, albeit through the presence of an additional light colored scalar. 

The remainder of the paper is structured as follows. In Section~\ref{htogammagamma} we discuss a possible explanation of the $h\to\gamma\gamma$ excess with colored scalar fields. We analyze the impact of all the relevant experimental inputs and specify the numerical procedure used to generate our results. We also provide predictions for $h\to Z\gamma$ and di-Higgs production using the best-case scenarios and associated parameters. The fields that can help accommodate  the $h\to\gamma\gamma$ excess are shown to appear naturally in matter unification scenarios in Section~\ref{color_scalars}. There we also correlate the lightness of these states with observable proton decay signatures. Finally, we conclude in Section~\ref{conclusions}.

\section{Enhanced di-photon signal with colored scalars}
\label{htogammagamma}
We study the influence of colored scalar fields coupling to the SM Higgs
doublet through $ \Phi^\dagger \Phi H^\dagger H$ interactions on Higgs production and decay signatures at the LHC. 
If sufficiently light, such states can significantly modify the loop-induced $gg\to h$, $gg\to hh$, $h \to \gamma Z$
and $h\to\gamma\gamma$ processes, while leaving other, tree-level dominated Higgs production and decay channels SM-like. 

After specifying the color and EW representations, each scenario ($\Phi$) can be parametrized in terms of the relevant Higgs couplings and masses of the weak components $\Phi_i$ of the multiplet. Different possible weak contractions of the $\Phi^{\dagger}\Phi H^{\dagger}H$ terms will in general induce different relative contributions to these couplings and masses. However, severe experimental constraints coming mainly from the $\rho$ parameter require an approximate custodial symmetry to be active in the EW symmetric scalar potential (c.f.~\cite{Barbieri:2004qk}).
It turns out that in this limit, one can without loss of generality consider a single interaction term of the form
\begin{equation}
\mathcal L_{} \ni -\lambda_\Phi (\Phi_{i a}^{\dagger}\Phi_{i a}) (H_j^{\dagger}H_j) =  -\lambda_{\phi} m_W \Phi_{i a}^\dagger \Phi_{i a} h + \ldots\,,
\label{eq:2Phi}
\end{equation}
where we have written out the summed over weak ($i,j$) and color ($a$) indices explicitly, and the dots denote further terms in the EW broken phase expansion of the Higgs fields. Furthermore, in the custodial limit, ${\Phi_i}$ are almost degenerate and we will use $m_{\phi}$ to denote their common mass.


Then, the partial decay width for $h\to\gamma\gamma$ at one loop is given by~\cite{SUSYHiggs,Chang:2012ta},
\begin{equation}
\label{eq:1}
\Gamma_{h\to\gamma\gamma}=\frac{G_{\mu}\alpha^{2}m_{h}^{3}}{128\sqrt{2}\pi^{3}}\left|\mathcal A_{1}({x}_{W})+\frac{4}{3}\mathcal A_{1/2}({x}_{t})+\underset{i}{\sum}\frac{\lambda_{\phi}}{g_{w}}\frac{m_{W}^{2}}{m_{\phi}^{2}}d(r_{\Phi})Q_{\Phi_i}^{2}\mathcal A_{0}({x}_{\phi})\right|^{2},
\end{equation}
where $G_\mu$ is the Fermi constant, $\alpha$ the fine structure constant, $g_w= \sqrt{4\pi \alpha}/\sin \theta_w$ and $\theta_w$ the Weinberg angle. Also, ${x}_{i}=m_{h}^{2}/(4m_{i}^{2})$ for $i=W,t,\phi$, 
while relevant one-loop functions are given by 
\begin{eqnarray}
\mathcal A_{1}({x})&=&-\left(2{x}^{2}+3{x}+3(2{x}-1)f({x})\right){x}^{-2},\\
\mathcal A_{1/2}({x})&=&2\left({x}+({x}-1)f({x})\right){x}^{-2},\\
\mathcal A_{0}({x})&=&-\left({x}-f({x})\right){x}^{-2},\\
f({x})&=&\left\{ \begin{array}{cc}
\arcsin^{2}\sqrt{{x}} & {x}\leq1\\
-\frac{1}{4}\left(\log\frac{1+\sqrt{1-{x}^{-1}}}{1-\sqrt{1-{x}^{-1}}}-i\pi\right)^{2} & {x}>1
\end{array}\right..
\end{eqnarray}
The first and the second term in Eq.~\eqref{eq:1} are the SM one-loop contributions from the $W$ and the
top quark, respectively. For $m_{h}=125\,\textrm{GeV}$, their numerical values are $\mathcal A_{1}({x}_{W})=-8.3$
and $\mathcal A_{1/2}({x}_{t})=1.4$. 
Finally, $d(r_{\Phi})$ is the
dimension of the color representation of $\Phi$, and $Q_{\Phi_i}$ the
electric charges of weak $\Phi_i$ components.

Analogously, the parton level $gg\to h$ cross section at partonic c.m.s.\ energy $\sqrt{\hat s}$ reads~\cite{SUSYHiggs,Chang:2012ta}
\[
\hat \sigma_{gg\to h}=\sigma_{0}m_{h}^{2}\delta(\hat{s}-m_{h}^{2}),
\]
\[
\sigma_{0}=\frac{G_{\mu}\alpha_{s}^{2}}{128\sqrt{2}\pi}\left|\frac{1}{2}\mathcal A_{1/2}({x}_{t})+\underset{i}{\sum}\frac{\lambda_{\phi}}{g_{w}}\frac{m_{W}^{2}}{m_{\phi}^{2}}C(r_{\Phi})\mathcal A_{0}({x}_{\phi})\right|^{2},
\]
where $\alpha_s$ is the strong coupling constant and $C(r_{\Phi})$ is the index of the color representation $r_{\Phi}$ of  $\Phi$. The only color representations we consider are triplet, sextet and octet as these are the ones that can contract directly with SM matter fields in $SU(3)$ space to yield a singlet.  We accordingly use $C(3)=1/2$, $C(6)=5/2$ and $C(8)=3$.

\subsection{Numerical Procedure}
Our goal is to confront available Higgs signal strength data with
possible contributions from light colored scalars. Parameters used to fit the data are $\lambda_{\phi}$
and $m_{\phi}$. Following~\cite{Espinosa:2012im} we define individual channel signal rates, normalized to their respective SM values as
\[
\mu_{i}=\frac{\left(\underset{j}{\sum}\sigma_{j\to h}\times \mathcal B_{h\to i}\right)}{\left(\underset{j}{\sum}\sigma_{j\to h}\times \mathcal B_{h\to i}\right)_{\mathrm{SM}}}\,,
\]
where the labels $j$ and $i$ denote the relevant Higgs production and detection channels. We furthermore denote the reported experimental values and variances  of $\mu_i$ by $\hat{\mu}_{i}$ and $\hat{\sigma}_{i}^{2}$, respectively  (listed in Table~\ref{tab:2}). 

A global $\chi^{2}$ is then defined as
\begin{equation}
\label{eq:5}
\chi^{2}(\lambda_{\phi},m_{\phi})=\underset{i}{{\sum}}\frac{(\mu_{i}(\lambda_{\phi},m_{\phi})-\hat{\mu}_{i})^{2}}{\sigma_{i}^{2}}\,,
\end{equation}
where we neglect correlations among the various terms as they are not supplied by
the experimental collaborations. As pointed out in~\cite{Giardino:2012dp},
theoretical uncertainties are only relevant for $\sigma_{gg\to h}^{\mathrm{SM}}$, where they amount to
a relative error of $\pm14\%$. To obtain the $\sigma_{i}$ values in Eq.~\eqref{eq:5}, we add this contribution in quadrature with the experimental errors ($\hat\sigma_i$) for each observable.
For each scenario ($\Phi$), we determine the minimum of the $\chi^{2}$ ($\chi_{min}^{2}$),
and define the 68\% ($1\sigma$) and 95\% ($2\sigma$)
best-fit regions as solutions to $\chi^{2}\leq \chi_{min}^{2}+\Delta\chi^{2}$, where $\Delta\chi^{2}$ are set by the appropriate cumulative distribution function.

\subsection{Data}
\label{data}
The relevant observables, which can be significantly affected by colored scalar contributions, are the recently measured LHC Higgs production rates in the $WW^{*}$, $ZZ^{*}$, $\gamma\gamma$
and $\gamma\gamma jj$ channels~\cite{key-6,key-7,key-14,key-16,key-17,key-18,key-19,key-20,key-15} listed in Table~\ref{tab:2}. In our fit we combine the independent measurements for each channel using weighted average.
The $\gamma\gamma jj$ channel is the only one with a significant contribution from vector boson fusion (VBF). The expected signal strength in this channel can be parametrized as~\cite{Espinosa:2012im}
\[
\mu_{\gamma\gamma jj}=\frac{0.033\,\sigma_{gg\to h}+\sigma^{}_{\mathrm{VBF}}}{0.033\,\sigma_{gg\to h}^{\mathrm{SM}}+\sigma_{\mathrm{VBF}}^{\mathrm{SM}}}\times \frac{\mathcal B_{h\to\gamma\gamma}}{\mathcal B^{\mathrm{SM}}_{h\to\gamma\gamma}},
\]
where $\sigma_{gg\to h}$ denotes the hadronic $gg\to h$ cross section and  the ratio $\sigma_{\mathrm{VBF}}^{\mathrm{SM}}/\sigma^{\mathrm{SM}}_{gg} \simeq 0.078$~\cite{Dittmaier:2011ti} remains almost constant when going from $7$\,TeV to $8$\,TeV c.m.s.\ energy at the LHC. Since VBF remains SM like in our scenarios, we can also identify $\sigma_{\mathrm{VBF}}^{\mathrm{}}\simeq \sigma_{\mathrm{VBF}}^{\mathrm{SM}}$.
All other channels are completely dominated by $gg\to h$ production alone. 

\begin{table}[htdp]
\begin{centering}
\begin{tabular}{|l|l|l|}
\hline 
CHANNEL & $\hat \mu_i \pm \hat\sigma_i$ & REFERENCE\tabularnewline
\hline \hline
pp$\rightarrow$ZZ{*}$\rightarrow$4l & $ $$1.25\pm0.55$ & ATLAS7+8~\cite{Giardino:2012dp}\tabularnewline
pp$\rightarrow$ZZ{*}$\rightarrow$4l & $0.85\pm0.3$ & CMS7+8~\cite{Giardino:2012dp}\tabularnewline
pp$\rightarrow$WW{*}$\rightarrow$4l & $1.4\pm0.5$ & ATLAS7+8~\cite{key-15}\tabularnewline
pp$\rightarrow$WW{*}$\rightarrow$4l & $0.7\pm0.4$ & CMS7+8~\cite{Giardino:2012dp}\tabularnewline
pp$\rightarrow$$\gamma\gamma$ & $1.7\pm0.55$ & ATLAS7+8~\cite{Giardino:2012dp}\tabularnewline
pp$\rightarrow$$\gamma\gamma$ & $1.5\pm0.4$ & CMS7+8~\cite{Giardino:2012dp}\tabularnewline
pp$\rightarrow$$\gamma\gamma jj$ & $3.2\pm0.9$ & CMS7~\cite{Giardino:2012dp}\tabularnewline
pp$\rightarrow$$\gamma\gamma jj$ & $1.6\pm0.8$ & CMS8~\cite{Giardino:2012dp}\tabularnewline
\hline 
\end{tabular}
\par\end{centering}
\caption{Data used in the analysis.}
\label{tab:2}
\end{table}

\subsection{Results}

We consider contributions of colored scalars listed in Table~\ref{tab:1} to Higgs production and decays. The fit parameters are the effective coupling $\lambda_{\phi}$ and the colored scalar mass $m_{\phi}$. In the SM reference scenario, $\chi_{min}^{2}/{\rm d.o.f.}=1.84$. 

\begin{table}[htdp]
\begin{centering}
\begin{tabular}{|l|l|l|}
\hline 
$SU(3)\times SU(2)\times U(1)$  & $\chi_{min}^{2}$\qquad{}\qquad{} & $\chi^{2}(ZZ,\, WW,\,\gamma\gamma,\,\gamma\gamma jj)$\tabularnewline
\hline \hline
$(\overline{\mathbf{3}},\mathbf{1},1/3)$  & $6.8$ & $(0.49,\,0.25,\,1.61,\,4.5)$\tabularnewline
$(\overline{\mathbf{3}},\mathbf{1},-2/3)$  & $7.2$ & $(0.22,\,0.09,\,2.20,\,4.7)$\tabularnewline
$(\mathbf{3},\mathbf{1},-4/3)$  & $6.1$ & $(0.31,\,0.30,\,2.36,\,3.1)$\tabularnewline
$(\mathbf{3},\mathbf{2},1/6)$  & $7.1$ & $(0.35,\,0.17,\,1.91,\,4.6)$\tabularnewline
$(\mathbf{3},\mathbf{2},7/6)$  & $6.5$ & $(0.20,\,0.21,\,2.55,\,3.5)$\tabularnewline
$(\mathbf{3},\mathbf{3},-1/3)$  & $7.4$ & $(0.04,\,0.01,\,2.62,\,4.7)$\tabularnewline
$(\overline{\mathbf{6}},\mathbf{1},-1/3)$  & $6.7$ & $(0.55,\,0.28,\,1.48,\,4.4)$\tabularnewline
$(\overline{\mathbf{6}},\mathbf{1},2/3)$  & $6.9$ & $(0.44,\,0.21,\,1.73,\,4.5)$\tabularnewline
$(\overline{\mathbf{6}},\mathbf{1},-4/3)$  & $7.4$ & $(0.07,\,0.03,\,2.56,\,4.7)$\tabularnewline
$(\overline{\mathbf{6}},\mathbf{3},-1/3)$  & \Green{$0.7$} & $(0.02,\,0.04,\,0.12,\,0.5)$\tabularnewline
$(\mathbf{8},\mathbf{2},1/2)$  & \Green{$1.3$} & $(0.03,\,0.00,\,0.01,\,1.2)$\tabularnewline
\hline 
SM & $7.4$ & $(0.04,\,0.01,\,2.63,\,4.7)$\tabularnewline
\hline 
\end{tabular}
\par\end{centering}
\caption{The list of colored scalars that couple to the SM fermions
at renormalizable level and corresponding $\chi_{min}^{2}$ from a fit to Higgs production and decay measurements. Last column shows contributions to $\chi_{min}^{2}$ from individual observables in the fit whereas the last row contains the SM result.}
\label{tab:1}
\end{table}

\begin{figure}
\begin{centering}
\includegraphics[scale=1.1]{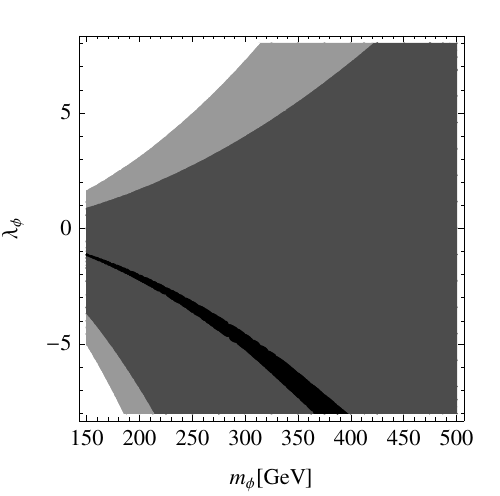}$\;\;\;\;\;$\includegraphics[scale=1.1]{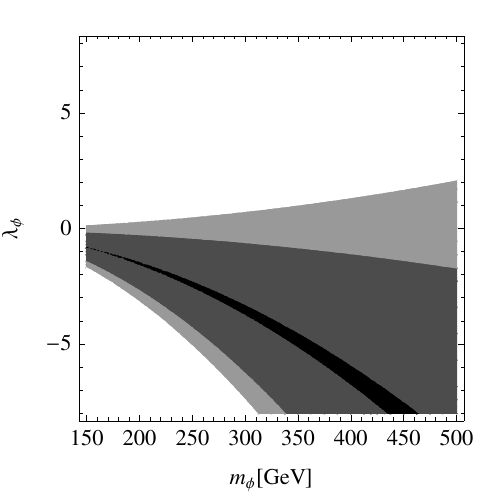}
\par\end{centering}
\caption{(left) $\chi^{2}$ plot for $(\mathbf{3},\mathbf{1},-4/3)$ state where $\chi_{min}^{2}/{\rm d.o.f.}=3.05$. Black strip is a region in parameter space
with the minimum $\chi^{2}$. Two other regions are the $1\sigma$ and $2\sigma$. (right) $\chi^{2}$ plot for $(\mathbf{3},\mathbf{1},8/3)$ state where
$\chi_{min}^{2}/{\rm d.o.f.}=0.35$.}
\label{fig:1}
\end{figure}

We notice two different behaviors in the interesting
range of parameters. Obviously, to enhance $h\to\gamma\gamma$,
a negative $\lambda_{\phi}$ is preferred, but this tends to lower $gg\to h$, affecting all the measured channels. The goal then would be to have as small $C(r_{\Phi})$
as possible and as big $d(r_{\Phi})Q_{\Phi_i}^{2}$ as possible, and
to have a small negative $\lambda_{\phi}$. It turns out, however, that none of the considered scalars
can accommodate all the data in this way. The best candidate from the list would be the $(\mathbf{3},\mathbf{1},-4/3)$
and we present its $\chi^{2}$ plot in the $(\lambda_{\phi},m_{\phi})$ plane in Fig.~\ref{fig:1}.
Clearly, $(\mathbf{3},\mathbf{1},-4/3)$ can fit $gg\to h$ well, but fails to enhance $h\to\gamma\gamma$. This is typical for most of the scalars listed in Table~\ref{tab:1}, as can be seen
from the last column which shows separate contributions for all observables in the fit. For the sake of the  argument, we also plot the scalar $(\mathbf{3},\mathbf{1},8/3)$ example, which fits
the data perfectly but does not couple directly to the SM fermions.

\begin{figure}
\begin{centering}
\includegraphics[scale=1.1]{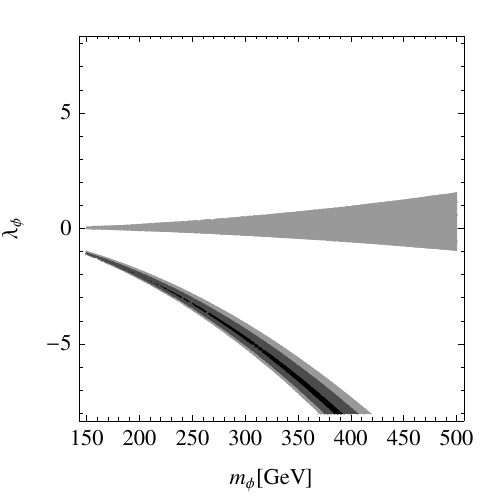}$\;\;\;\;\;$\includegraphics[scale=1.1]{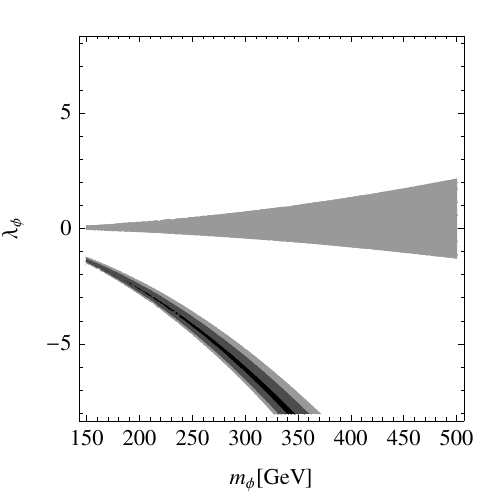}
\par\end{centering}
\caption{(left) $\chi^{2}$ plot for $(\overline{\mathbf{6}},\mathbf{3},-1/3)$ state contribution where $\chi_{min}^{2}/{\rm d.o.f.}=0.35$. (right) $\chi^{2}$ plot for $(\mathbf{8},\mathbf{2},1/2)$ state contribution where
$\chi_{min}^{2}/{\rm d.o.f.}=0.63$.}
\label{fig:2}
\end{figure}

Another possibility is to allow $\lambda_{\phi}C(r_{\Phi})$ to be negative
and large, so that contribution of the scalar in the loop is twice
the contribution of the top quark but with opposite sign. 
In this case $gg\to h$ will again have the same value as in the SM. This particular scenario 
to keep $gg\to h$ at the SM level has been observed in Refs.~\cite{Dobrescu:2011aa,Kribs:2012kz}. To accomplish this, we need large $C(r_{\Phi})$ since $\lambda_{\phi}$ is constrained by perturbativity arguments.
It turns out that we have two good candidates listed in Table~\ref{tab:1} that can accomplish the task. These are the color sextet $(\overline{\mathbf{6}},\mathbf{3},-1/3)$ and color octet
$(\mathbf{8},\mathbf{2},1/2)$. The corresponding $\chi^{2}$ plots for these states are presented in Fig.~\ref{fig:2}. Note that the optimal parameter space in both
cases is very narrow. This practically fixes the allowed $\lambda_\phi$ for a given value of $m_\phi$ and vice versa.   We obtain $\chi_{min}^{2}/{\rm d.o.f.} = 0.35$ and $\chi_{min}^{2}/{\rm d.o.f.} = 0.63$ for 
$(\overline{\mathbf{6}},\mathbf{3},-1/3)$ and $(\mathbf{8},\mathbf{2},1/2)$ scenarios, respectively. 

\subsection{Other Constraints}

In this best-fit region, the sextet (octet) masses above $380$\,GeV ($340$\,GeV) quickly lead to non-perturbative values of the coupling $\lambda_\Phi>\sqrt{4\pi}$. On the other hand, the remaining interesting range of masses is not yet completely excluded by direct searches at the LHC as recently emphasized in~\cite{Altmannshofer}. In particular the mass window between $200$~GeV and $320$~GeV is allowed by present searches for colored scalars at the LHC. Furthermore, current experimental analyses assume that such scalars are narrow and decay $100\%$ to pairs of jets, none of which is required or predicted by our fit to the Higgs data (for a recent discussion on light colored resonances escaping present experimental searches c.f.~\cite{colored}). 

Another important issue to address in these scenarios, given the large required $\lambda_\phi$ couplings, concerns vacuum stability and perturbativity of the scalar potential. We focus here on the $(\mathbf{8},\mathbf{2},1/2)$ state for concreteness, while similar conclusions hold for the $(\overline{\mathbf{6}},\mathbf{3},-1/3)$ as well.  A recent study of vacuum decay constraints on colored scalars coupling to the Higgs~\cite{Reece:2012gi} included the $(\mathbf{8},\mathbf{2},1/2)$ representation and found that in the low energy effective theory comprising the SM and the color octet weak doublet state, and for the interesting range of color octet masses as extracted from our analysis, vacuum (meta)stability constraints  can be satisfied, provided a quartic term of the form 
\begin{equation}
\mathcal L \ni -\lambda^{}_{4\Phi} {\rm }(\Phi_{a i}^\dagger  \Phi_{a i})^2\,,
\label{eq:4Phi}
\end{equation}
 is present and the couplings satisfy $\lambda_{4\Phi} \gtrsim \lambda_\Phi^2/8 \lambda$, where $\lambda$ is the SM Higgs quartic coupling $\lambda \equiv 2 m_h^2/v^2_{\rm EW}$ and $v_{\rm EW} \simeq 246$~GeV  is the EW condensate. While this shows that all involved quartic couplings can be perturbative at the EW scale, this is not necessarily true up to arbitrary high scales ($\mu$). We study the issue using renormalization group equations (RGEs). For concreteness we consider the two relevant $\Phi$ quartics in Eqs.~\eqref{eq:2Phi} and~\eqref{eq:4Phi}.\footnote{For a full list of possible quartics involving the $({\bf 8},{\bf 2},1/2)$ state and the SM Higgs c.f.~\cite{Manohar:2006ga}.}  Taking into account also the effects of the top yukawa ($Y_t$) interaction, the RGEs for $\lambda$, $\lambda^{}_\Phi$  and $\lambda^{}_{4\Phi}$ at one loop are given by
\begin{align}
16\pi^{2}\frac{d\lambda_{}}{d\ln\mu}&=24\lambda_{}^{2}+16\lambda_{\Phi}^{2}+12Y_{t}^{2}\lambda_{}-6Y_{t}^{4}\,, \nonumber\\
16\pi^{2}\frac{d\lambda^{ }_{\Phi}}{d\ln\mu}&=4\lambda_{\Phi}^{ 2}+12\lambda_{}\lambda^{ }_{\Phi}+68\lambda^{ }_{4\Phi}\lambda_{\Phi}+6Y_{t}^{2}\lambda^{ }_{\Phi}\,,\nonumber\\
16\pi^{2}\frac{d\lambda^{ }_{4\Phi}}{d\ln\mu}&=80\lambda^{ 2}_{4\Phi}+2\lambda_{\Phi}^{ 2}\,.
\end{align}
We observe that the positive beta function of $\lambda^{ }_{4\Phi}$ necessarily drives this coupling to large positive values at high scales. In particular, for the parameter range preferred by the fit to the Higgs data and satisfying the vacuum metastability constraint, $\lambda^{ }_{4\Phi}$ develops a Landau pole already at  scales $\mu_{\rm nonpert.} \lesssim 10$~TeV. Such a scenario can thus only represent a low energy effective theory, which needs to be extended below $\mu_{\rm nonpert.}$. One possibility without introducing new light degrees of freedom is to utilize the couplings of $\Phi$ to matter fields. In particular, couplings to quarks of the form $Y_{q\Phi} \bar Q_L \Phi q_R$ can induce negative contributions to the beta functions of  $\lambda_{\Phi}$ and $\lambda_{4\Phi}$ proportional to $Y_t^2 Y_{q\Phi}^2$  and $Y_{q\Phi}^4$, respectively. Here, $Q_L$, $q_R$ refer to left- and right-handed quark fields, respectively. If large enough, such contributions can, in principle, stabilize the color octet quartics. However, $Y_{q\Phi}$  also induce mixing among the various possible terms in the most general scalar potential involving the $({\bf 8},{\bf 2},1/2)$ state~\cite{Manohar:2006ga} and its complete RGE study is clearly beyond the scope of the present paper. Finally we note that the presence of $Y_{q\Phi}$ might have interesting phenomenological consequences in both flavor observables as well as for collider signatures of the color octet scalar (c.f.~\cite{Manohar:2006ga,Altmannshofer,Gerbush:2007fe}). A detailed study of the implications of sizable $Y_{q\Phi}$ on the UV behavior of the $({\bf 8},{\bf 2},1/2)$ scalar potential and the associated phenomenology is in progress.

\subsection{Predictions for $h\to Z\gamma$ and di-Higgs production }

In the SM, the $h\to Z\gamma$ decay is  generated at the loop level in a very similar
way as $h\to\gamma\gamma$. Being extremely suppressed, this channel has not yet been explored by the experimental collaborations. However, once it is measured,
it will be an extremely useful test of possible explanations of the $h\to\gamma\gamma$
anomaly since, in general, these two rates are related. 
The partial decay width for $h\to Z\gamma$ including colored scalars contribution is given by~\cite{Carena:2012xa}
\begin{eqnarray}
&&\Gamma_{h\to Z\gamma} = \frac{G_F^2 m_W^2 \alpha}{64\pi^4} m_h^3\left( 1-\frac{m_Z^2}{m_h^2} \right)^3\nonumber\\
&&\times \left|{\cos\theta_{W}\mathcal C_{1}({x}^{-1}_{W},y_{W})+\frac{2(1-(8/3)\sin^{2}\theta_{W})}{\cos\theta_{W}}\mathcal C_{1/2}({x}^{-1}_{t},y_{t})}+{\frac{v_{\rm EW}\sin\theta_{W}}{2} \sum_i{\lambda_{\phi}m_{W}g_{Z\Phi_i \Phi_i}d(r_{\Phi})Q_{\Phi_i}}{m_{\phi}^{-2}}\mathcal C_{0}({x}^{-1}_{\phi},y_{\phi})}\right|^{2}\nonumber\\
\label{hzgamma}
\end{eqnarray}
where $y_{i}\equiv{4m_{i}^{2}}/{m_{Z}^{2}}$ and the coupling of
$\phi$ to $Z$ boson is given in units of $|e|$, that is $g_{Z\Phi_i\Phi_i}=2({T_{\Phi_i}^{3}-Q_{\Phi_i}\sin^{2}\theta_{W}})/{\sin2\theta_{W}}$. Here  $T_{\Phi_i}^{3}$ represents the value of the weak isospin of $\Phi_i$ and in Eq.~\eqref{hzgamma} we sum over all $i$ within the given weak multiplet $\Phi$.  
The relevant one-loop functions are defined as \[
\mathcal C_{0}(x,y)=I_{1}(x,y),\]
\[
\mathcal C_{1}(x,y)=4(3-\tan^{2}\theta_{W})I_{2}(x,y)+\left((1+2x^{-1})\tan^{2}\theta_{W}-(5+2x^{-1})\right)I_{1}(x,y),\]
\[
\mathcal C_{1/2}(x,y)=I_{1}(x,y)-I_{2}(x,y),\]
where\[
I_{1}(x,y)=\frac{x y}{2(x-y)}+\frac{x^{2} y^{2}}{2(x-y)^{2}}\left(f(x^{-1})-f(y^{-1})\right)+\frac{x^{2}y}{(x-y)^{2}}\left(g(x^{-1})-g(y^{-1})\right),\]
\[
I_{2}(x,y)=-\frac{x y}{2(x-y)}\left(f(x^{-1})-f(y^{-1})\right),\]
\[
g(x)=\sqrt{x^{-1}-1}\arcsin\sqrt{x},\;\;\;\;x\geq1.\]
The SM contributions to $h \to Z \gamma$  induced by the $W$ boson  and the top quark are proportional to $\cos\theta_{W}\mathcal C_{1}(x^{-1}_{W},y_{W})=5.8$ and  ${2(1-(8/3)\sin^{2}\theta_{W})}\mathcal C_{1/2}(x^{-1}_{t},y_{t})/{\cos\theta_{W}}=-0.3$, respectively. Clearly, the SM result is dominated by the $W$ boson contribution.

We calculate 
$
\mu_{Z\gamma} \equiv {\Gamma_{h\to Z \gamma}}/{\Gamma_{h\to Z\gamma}^{\rm SM}}\,,
$
for the best-fit values of
coupling $\lambda_{\phi}$ and mass $m_{\phi}$ for the scalars
that can accommodate current Higgs data. For the $(\mathbf{8},\mathbf{2},1/2)$
state 
we predict $\mu_{Z\gamma}=(0.90\pm0.01)$ 
throughout the interesting range of scalar masses and couplings.
Similarly, for the $(\mathbf{\overline{6}},\mathbf{3},-1/3)$ state
we get $\mu_{Z\gamma}=(0.70\pm0.02)$, again independent of the scalar mass in the interesting range of parameters.
We observe that both scenarios predict a mild suppression of the $h\to Z\gamma$ decay rate and that a $20\%$ measurement of the relevant branching fraction could start probing the scalar color sextet explanation of the $h\to \gamma\gamma$ excess, something possibly in reach of the 14\,TeV LHC with several hundred ${\rm fb}^{-1}$~\cite{Gainer:2011aa}.

Another interesting related process is the di-Higgs production via gluon fusion, i.e., $gg\to hh$, which is again loop suppressed in the SM and thus potentially sensitive to non standard contributions. We accordingly turn our attention to this process. The di-Higgs production can be significantly affected in the presence of light colored fields. This is especially true for the regime we explore where $|\lambda_{\phi}|$ is large ($\lambda_{\phi}<0$) and colored state mass $m_{\phi}$ is relatively small~\cite{Kribs:2012kz}. We accordingly evaluate $\mu_{hh} \equiv {\sigma_{gg\to hh}}/{\sigma_{gg\to hh}^{\rm SM}}$ for the best-fit values of coupling $\lambda_{\phi}$ and mass $m_{\phi}$ for both scalars that accommodate current Higgs data in a satisfactory manner. We use the parton level loop-induced top quark contribution towards $gg\to hh$ amplitude~\cite{Glover:1987nx} as well as the loop-induced colored scalar contribution~\cite{Kribs:2012kz} to evaluate total cross-sections $\sigma_{gg \to hh}^{\mathrm{SM}}$ and $\sigma_{gg \to hh}$ that are relevant for the current LHC energy reach. We integrate the parton level $gg\to hh$ cross sections using the LO MSTW2008 parton distribution functions~\cite{Martin:2009iq} with fixed factorization and normalization scales of $2 m_h$.  

At the 8\,TeV c.m.s.\ energy LHC we obtain for the $(\mathbf{8},\mathbf{2},1/2)$ and $(\mathbf{\overline{6}},\mathbf{3},-1/3)$ states an enhancement of $\mu_{hh}=(200\pm 60)$ and $\mu_{hh}=(140\pm 40)$ respectively, for the best fit regions of couplings and almost independent of the scalar masses in their interesting range. We have checked that these values also hold for the 14\,TeV LHC energy. 
 We observe that both scenarios predict significant enhancement of the total cross-section for di-Higgs production via gluon fusion with respect to the SM predictions. The signals and discovery strategies within the regime of enhanced di-Higgs production at LHC have been discussed extensively in Ref.~\cite{Kribs:2012kz}, where the relevant rates for various final states can be found for both 8\,TeV and 14\,TeV c.m.s.\ energy LHC.

\section{Colored scalars and matter unification}
\label{color_scalars}
Our analysis singles out two particular colored scalars---$(\mathbf{8},\mathbf{2},1/2)$ and $(\overline{\mathbf{6}},\mathbf{3},-1/3)$---as potential candidates that can consistently address the observed enhancement in $h\to\gamma\gamma$ decay channel if light enough. Here we want to comment on matter unification scenarios that predict these scalars to be in a required mass range. Note that although both scalars couple directly to matter they do not mediate proton decay. 

\subsection{Color octet}

Color octet state is an appealing source of new physics. For example, it is the only scalar beside the Higgs doublet that can be consistently coupled to the SM quarks within the Minimal Flavor Violation framework~\cite{Manohar:2006ga}. Octet production at the LHC and relevant electroweak constraint on its mass and Yukawa couplings in that particular context have been extensively studied~\cite{Gresham:2007ri}. More recently, there have been numerous studies of the color octet influences on Higgs physics in view of LHC data and potential signals~\cite{Arnold:2011ra,Bai:2011aa,Dobrescu:2011aa,Kumar:2012ww,Kribs:2012kz}. Note that in all these instances the presence of light octet state is simply assumed. 

There are, however, viable unifying models that predict existence of a light color octet state and correlate it with proton decay signatures~\cite{Perez:2007rm,Dorsner:2006dj,Dorsner:2007fy,Bertolini:2012im}. We discuss two particular models based on the $SU(5)$ gauge group~\cite{Georgi:1974sy} in what follows. The first model~\cite{Perez:2007rm} uses one $5$-dimensional ($\mathbf{5}$) and one $45$-dimensional ($\mathbf{45}$) scalar representation to accommodate charged fermion masses~\cite{Georgi:1979df} and a set of extra fermion fields from one $24$-dimensional  ($\mathbf{24}_F$) representation~\cite{Bajc:2006ia} to accommodate neutrino masses. The phenomenology of light octet state that resides in the $45$-dimensional scalar representation within this particular context has been addressed in Ref.~\cite{Perez:2008ry}. The second model uses three scalar representations---$\mathbf{5}$, $\mathbf{15}$ and $\mathbf{45}$---to accommodate all fermion masses~\cite{Dorsner:2006dj,Dorsner:2007fy}. These two models, beside the usual matter representations, also use one $24$-dimensional scalar representation ($\mathbf{24}$) to break $SU(5)$ down to $SU(3) \times SU(2) \times U(1)$.

In both scenarios an upper bound on octet mass is correlated with the observable partial proton decay lifetimes. 
In view of the latest experimental data on partial proton decays modes~\cite{Miura:2010zz,:2012rv}, in particular on $p \rightarrow \pi^0 e^+$, we update these predictions for both models to demonstrate this correlation. 

\begin{figure}[t]
\begin{center}
\includegraphics[width=7cm]{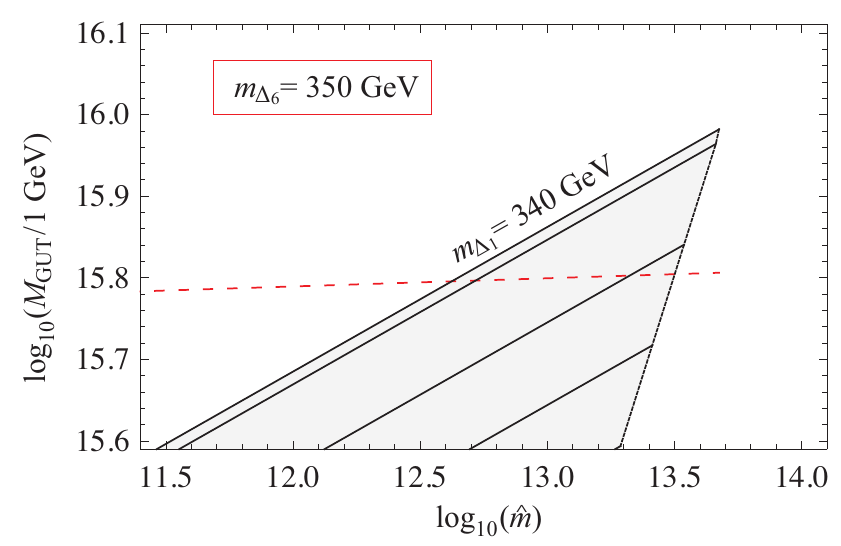}
\end{center}
\caption{Viable unification for the model with $5$-dimensional and $45$-dimensional scalar representations and an extra set of fermions in $24$-dimensional representation in $\hat{m}$ vs.\ $M_{\mathrm{GUT}}$ plane at the one-loop level for central values of low-energy observables. Solid lines, going from top to bottom, correspond to $m_{\Delta_1}=340$\,GeV, $m_{\Delta_1}=500$\,GeV, $m_{\Delta_1}=5$\,TeV and $m_{\Delta_1}=50$\,TeV. Horizontal dashed line is due to the constraint imposed by experimental limit on partial proton decay lifetime through $p \to \pi^0 e^+$ on $m_{\Delta_1}=340$\,GeV case.}
\label{fig:3}
\end{figure}

We show in Fig.~\ref{fig:3} a viable unification for the model with one $5$-dimensional and one $45$-dimensional scalar representation and one $24$-dimensional matter representation. 
Numerical procedure that is used to establish unification of gauge couplings and implement proton decay constraints is described in detail in Ref.~\cite{Dorsner:2009mq}. Here, we outline the most important points that lead to results presented in Fig.~\ref{fig:3}. The scalar fields are denoted as $\mathbf{5} \equiv (\Psi_D,
\Psi_T) = (\mathbf{1},\mathbf{2},1/2)\oplus(\mathbf{3},\mathbf{1},-1/3)$, $\mathbf{45}\equiv(\Delta_1, \Delta_2, \Delta_3, \Delta_4, \Delta_5,
\Delta_6, \Delta_7) = (\mathbf{8},\mathbf{2},1/2)\oplus
(\overline{\mathbf{6}},\mathbf{1}, -1/3) \oplus (\mathbf{3},\mathbf{3},-1/3)
\oplus (\overline{\mathbf{3}}, \mathbf{2}, -7/6) \oplus (\mathbf{3},\mathbf{1},
-1/3) \oplus (\overline{\mathbf{3}}, \mathbf{1}, 4/3) \oplus (\mathbf{1},
\mathbf{2}, 1/2)$ and $\mathbf{24}\equiv(\Sigma_8
, \Sigma_3, \Sigma_{(3,2)}, \Sigma_{(\overline{3},2)},
\Sigma_{24}) = (\mathbf{8},\mathbf{1},0)\oplus(\mathbf{1},\mathbf{3},0)
\oplus(\mathbf{3},\mathbf{2},-5/6)\oplus(\overline{\mathbf{3}},\mathbf{2},5/6)
\oplus(\mathbf{1},\mathbf{1},0)$. The extra fermions in $\mathbf{24}_F \equiv (\rho_8,\rho_3, \rho_{(3,2)}, \rho_{(\bar{3}, 2)},
\rho_{24})=(\mathbf{8},\mathbf{1},0)\oplus(\mathbf{1},\mathbf{3},0)\oplus(\mathbf{3},\mathbf{2},-5/6)
\oplus(\overline{\mathbf{3}},\mathbf{2},5/6)\oplus(\mathbf{1},\mathbf{1},0)$ are related through a following set of mass relations~\cite{Perez:2007rm}
\begin{equation}
\label{condition}
m_{\rho_8}=\hat{m}m_{\rho_3}, \qquad m_{\rho_{(3,2)}}=m_{\rho_{(\bar{3}, 2)}}=\frac{(m_{\rho_3}+m_{\rho_8})}{2}.
\end{equation}
Here, $\hat{m}$ is a dimensionless free parameter that describes the mass splitting between masses of $\rho_8$ and $\rho_3$. We accordingly present our findings in a $\hat{m}$ vs.\ $M_{\mathrm{GUT}}$ plane, where $M_{\mathrm{GUT}}$ represents the scale of gauge coupling unification. $M_{\mathrm{GUT}}$ is maximized through numerical procedure that varies scalar and fermion masses, in accordance with mass splitting constraints of Eq.~\eqref{condition}, in the following ranges: $200$\,GeV$\leq m_{\Sigma_3},  m_{\Delta_1}, m_{\Delta_2}, m_{\Delta_4}, m_{\Delta_7}, m_{\rho_3}, m_{\rho_8}, m_{\rho_{(3,2)}}, m_{\rho_{(\bar{3}, 2)}} \leq M_{\mathrm{GUT}}$, $10^{12}$\,GeV$\leq m_{\Psi_T}, m_{\Delta_3}, m_{\Delta_5} \leq M_{\mathrm{GUT}}$ and $10^5$\,GeV$\leq m_{\Sigma_8} \leq M_{\mathrm{GUT}}$~\cite{Bajc:2006ia}. 

Solid lines in Fig.~\ref{fig:3}, going from top to bottom, correspond to $m_{\Delta_1}=340$\,GeV, $m_{\Delta_1}=500$\,GeV, $m_{\Delta_1}=5$\,TeV and $m_{\Delta_1}=50$\,TeV. Horizontal dashed line is due to a constraint imposed by experimental results on proton decay through $p \to \pi^0 e^+$ on unification case for $m_{\Delta_1}=340$\,GeV. Note that the difference between constraints on $m_{\Delta_1}=340$\,GeV case and $m_{\Delta_1}=50$\,TeV case is practically negligible as it is dominated by the difference in values of the appropriate unified gauge coupling at the GUT scale for these two cases. 

The most important point is that all unification scenarios below the dashed line in Fig.~\ref{fig:3} are excluded by experimental limits on $p \to \pi^0 e^+$. The proton decay signature through $p \to \pi^0 e^+$ channel is derived assuming that the Yukawa matrices for matter fields are symmetric. This assumption allows for light $\Delta_6$ as it prevents $\Delta_6$ to couple to a quark-quark pair~\cite{Dorsner:2012nq}. That, on the other hand, renders $\Delta_6$ innocuous as far as proton decay constraints are concerned. The same assumption removes dependence on unitary redefinitions of quark and lepton fields from proton decay operators induced through tree-level exchange of heavy gauge bosons. Note that the mass of $\Delta_6 = (\overline{\mathbf{3}}, \mathbf{1}, 4/3)$ needs to be below $560$\,GeV if it is to explain the $g-2$ anomaly of muon through perturbative Yukawa couplings~\cite{Dorsner:2011ai}. We take it to be $m_{\Delta_6}=350$\,GeV to generate Fig.~\ref{fig:3}. If $\Delta_6$ mass is closer to $560$\,GeV the allowed GUT scale would be slightly raised with respect to what is shown in Fig.~\ref{fig:3}. One should also worry that such a light colored state that resides in the same representation of $SU(5)$ as the octet field could spoil results for the satisfactory enhancement of $h \to \gamma \gamma$ signal. For example, if $(\overline{\mathbf{3}}, \mathbf{1}, 4/3)$ is to couple to the Higgs field with the same strength as the octet components one would need to simultaneous fit data with both fields to address the viability of such scenario. In the most general case one would have a situation where the triplet and the octet have different couplings to the Higgs boson and different masses. 

To generate results shown in Fig.~\ref{fig:3} we update some of input parameters with regard to what is used in Ref.~\cite{Dorsner:2009mq} to produce partial decay width for $p \to \pi^0 e^+$. We use $\alpha_s(m_Z)=0.1184$~\cite{PDG2012}, ${\tau}_{p \rightarrow \pi^0 e^+}>  1.3 \times 10^{34}$\,years~\cite{Miura:2010zz,:2012rv} and $\hat \alpha=-0.0112$\,GeV$^3$~\cite{Aoki:2008ku}. 
Here $\hat \alpha$ is the relevant nucleon matrix element.
The predicted proton lifetime for $p \rightarrow \pi^0 e^+$ due to gauge mediation is at most a factor of 5 above the current experimental limit for the $m_{\Delta_1}=340$\,GeV case while the proton lifetime due to scalar mediation is already at the present limit.

The GUT scale in Fig.~\ref{fig:3} is maximized by imposing a lower bound on proton decay mediating scalars, i.e., ${\Psi_T}$, ${\Delta_3}$ and ${\Delta_5}$, to be $10^{12}$\,GeV. It has been recently shown~\cite{Dorsner:2012nq} that the scalar exchange dominated proton decay in the models with $5$- and $45$-dimensional scalar representations with symmetric Yukawa couplings to matter fields constrains the mass of ${\Psi_T}$ through experimental data on $p \rightarrow K^+ \bar{\nu}$ channel to be  above $1.2 \times 10^{13}  (100\,\mathrm{GeV}/v_5)$ and $1.5 \times 10^{11} (100\,\mathrm{GeV}/v_5)\,\mathrm{GeV}$ in the most and least conservative case, respectively. We use ${\tau}_{p \rightarrow K^+ \bar{\nu}}>4.0 \times 10^{33}$\,years~\cite{Miura:2010zz}, while the VEVs for $5$- and $45$-dimensional representations---$v_5$ and $v_{45}$---satisfy $|v_5|^2/2+12 |v_{45}|^2=v_{\rm EW}^2$. This conclusively shows that $10^{12}$\,GeV is a reliable lower bound on the mass of proton decay mediating scalars for one-loop unification considerations.   

\begin{figure}[t]
\begin{center}
\includegraphics[width=7cm]{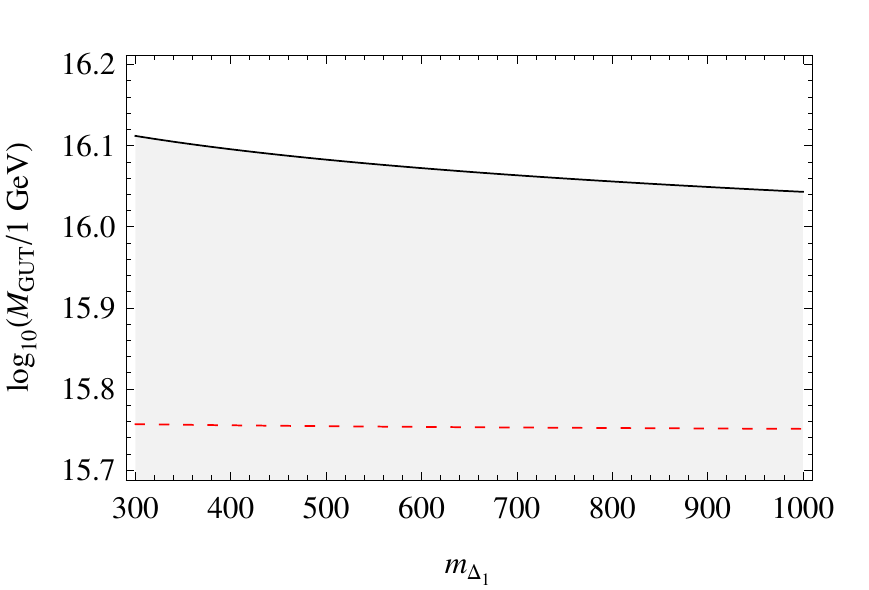}
\end{center}
\caption{Viable unification for the model with $5$-, $15$- and $45$-dimensional scalar representations in $m_{\Delta_1}$ vs.\ $M_{\mathrm{GUT}}$ plane at the one-loop level for central values of low-energy observables. Horizontal dashed line represents a limit due to the constraint imposed by experimental limit on partial proton decay lifetime through $p \to \pi^0 e^+$.}
\label{fig:4}
\end{figure}

We present in Fig.~\ref{fig:4} a viable unification for the model with $5$-, $15$- and $45$-dimensional scalar representations in a $m_{\Delta_1}$ vs.\ $M_{\mathrm{GUT}}$ plane at the one-loop level for central values of low-energy observables. We use $200$\,GeV$\leq m_{\Sigma_3}, m_{\Delta_1}, m_{\Delta_2}, m_{\Delta_4}, m_{\Delta_6}, m_{\Delta_7}, m_{\rho_{(3,2)}}, m_{\rho_{(\bar{3}, 2)}}, m_{\Phi_a}, m_{\Phi_c} \leq M_{\mathrm{GUT}}$, $10^{12}$\,GeV$\leq m_{\Psi_T}, m_{\Delta_3}, m_{\Delta_5}, m_{\Phi_b} \leq M_{\mathrm{GUT}}$, $10^5$\,GeV$\leq m_{\Sigma_8} \leq M_{\mathrm{GUT}}$, where $\mathbf{15}= (\Phi_a, \Phi_b, \Phi_c) = (\mathbf{1},\mathbf{3},1)\oplus
(\mathbf{3},\mathbf{2},1/6)\oplus(\mathbf{6},\mathbf{1},-2/3)$. In this case the predicted proton partial lifetime for $p \rightarrow \pi^0 e^+$ channel due to gauge mediation is at most a factor of 26 above the current experimental limit for $m_{\Delta_1}=300$\,GeV. The proton lifetime due to scalar mediation, on the other hand, is at the present limit.

The relevant coupling of the octet to the-would-be Higgs field $h$ originates from the following set of $SU(5)$ contractions: $ \lambda_1 \mathbf{5}^*_{\alpha} \mathbf{5}^{\alpha} \mathbf{45}_{\delta}^{\beta \gamma}  
\mathbf{45}^{*\,\delta}_{\beta \gamma}$ and $\lambda_2 \mathbf{5}^*_\alpha \mathbf{5}^\beta \mathbf{45}_{\gamma}^{\alpha \delta}  
\mathbf{45}_{\beta \delta}^{*\,\gamma}$. The couplings of the neutral  and charged component of the octet $\Delta_1$ to the-would-be SM Higgs, under the assumption that the SM doublet primarily originates from $5$-dimensional representation, are $\lambda_{\phi^0}=2 \lambda_1+\lambda_2$ and $\lambda_{\phi^+}=2 \lambda_1$, respectively. Any mixing between the doublets in $5$- and $45$-dimensional representation can be easily accounted for. In any case, we need to go to the limit $v_5 > v_{45}$. To reproduce the setup used in Section~\ref{htogammagamma} where the $h\to\gamma\gamma$ excess is accounted for via custodial symmetric color octet loops one needs to assume that $\lambda_2$ is much smaller than $\lambda_1$.

\subsection{Color sextet}
The color sextet $(\overline{\mathbf{6}}, \mathbf{3}, -1/3)$ resides in $50$- and $70$-dimensional representations of $SU(5)$~\cite{Slansky:1981yr}. It has been shown that it can provide for gauge coupling unification within an $SU(5)$ framework~\cite{Stone:2011dn}. It is, however, difficult to connect its lightness to proton decay or correlate its Yukawa couplings with the origin of masses for matter fields without additional assumptions in that particular setup. This seems, instead, to require an $SO(10)$ embedding scheme. Namely, the color sextet is part of $126$- and $210$-dimensional representations of $SO(10)$. Both representations are frequently used in model building with the former one being crucial in explaining observed fermion masses. For example, colored scalars, which are part of $126$-dimensional representation of $SO(10)$ and are light enough to be accessible at the LHC, have recently been proposed in Refs.~\cite{Bertolini:2012im,Babu:2012pp,Babu:2012iv}. In fact, there already exists a viable setup with an intermediate-scale sextet~\cite{Bertolini:2012im}. It might thus be feasible to have very light $(\overline{\mathbf{6}}, \mathbf{3}, -1/3)$ scalar that would originate from an $SO(10)$ model. We leave it to future studies.  

\section{Conclusions}
\label{conclusions}
We have considered the impact of light colored scalars that can couple directly to matter fields on the recently observed  $h\to\gamma\gamma$ excess. We find two viable scenarios where two states---$(\mathbf{8},\mathbf{2},1/2)$ and $(\overline{\mathbf{6}},\mathbf{3},-1/3)$---can individually influence the excess in a constructive way and remain in excellent agreement with all available data. The colored states in question should have a substantial coupling to the SM Higgs of order one and a mass of order $300$\,GeV (or below) in order to explain the data. In particular, perturbativity arguments require that the sextet and octet masses should be below $380$\,GeV and $340$\,GeV, respectively. 
The best fit values for the colored scalar masses and couplings to the Higgs are used to generate predictions for $h\to Z\gamma$ and di-Higgs production. 
We find moderate suppression of the partial decay width $h\to Z\gamma$ with regard to the SM value. The di-Higgs production, on the other hand, is enhanced by at least a factor of a hundred with respect to the SM prediction at both the 8~TeV and 14~TeV LHC. We subsequently study extensions of the SM where these states naturally appear with primary focus on matter unification models based on $SU(5)$ gauge group. It is shown that two simple models correlate light color octet mass with observable proton decay. For the color octet mass of $340$\,GeV the predicted partial proton decay lifetime through $p \rightarrow \pi^0 e^+$ channel is a factor of 5 (25) above the current limit for the model with extra fermions (scalars) in $24$-dimensional ($15$-dimensional) representation. In the model with extra fermions it is also possible to accommodate the $(g-2)_\mu$ anomaly, albeit through a presence of an additional light colored scalar. In conclusion, the presented setup relates high scale matter unification and matter stability to Higgs physics via the effects of light colored scalar states.   

\begin{acknowledgments}
This work was supported in part by the Slovenian Research Agency. I.D.\ acknowledges support by SNSF through the SCOPES project IZ74Z0\_137346.
\end{acknowledgments}

\end{document}